\documentclass[aps,prb,twocolumn,showpacs,final,a4paper,superscriptaddress,nofootinbib]{revtex4-2}

\usepackage{float}
\usepackage[dvips,final]{graphicx}
\usepackage{dcolumn}
\usepackage{bm}
\usepackage{textcomp}
\usepackage{amsmath,amssymb,amsfonts}
\usepackage{color}
\usepackage{hyperref} 
\hypersetup{colorlinks,linkcolor={blue},citecolor={blue},urlcolor={blue}}  
\usepackage{lipsum}
\usepackage{mleftright} 
\mleftright
\usepackage[english]{babel}
\makeatletter
\let\exp@select@language\selectlanguage
\def\selectlanguage#1{\exp@select@language{english}}
\makeatother

\makeatletter
\newcommand{\ie}{\emph{i.e.}\@ifnextchar.{\!\@gobble}{}}
\makeatother

\begin{document}

\title{Influence of Markovianity and self-consistency on time-resolved spectral functions of driven quantum systems }

\author{Thomas Blommel}
\affiliation{Department of Chemistry and Biochemistry, University of California, Santa Barbara, California, USA}
\affiliation{Materials Department, University of California, Santa Barbara, California, USA}

\author{M. Rey Lambert\textsuperscript{*}}
\affiliation{Department of Physics, University of California, Santa Barbara, California, USA}

\author{Michael A. Kurniawan\textsuperscript{*}}
\affiliation{Materials Department, University of California, Santa Barbara, California, USA}

\author{
Annabelle Canestraight}
\affiliation{Department of Chemical Engineering, University of California, Santa Barbara, California, USA}

\author{Vojtech Vlcek}
\affiliation{Department of Chemistry and Biochemistry, University of California, Santa Barbara, California, USA}
\affiliation{Materials Department, University of California, Santa Barbara, California, USA}

\date{\today}%
\begin{abstract}
We present a systematic comparison of the real-time Dyson expansion (RTDE) with established non-equilibrium Green's function approaches for simulating driven, interacting quantum systems. Focusing on density matrix dynamics, time-off-diagonal Green’s functions, and time-resolved photoemission spectra, we benchmark RTDE against fully self-consistent Kadanoff–Baym equation (KBE) calculations, the generalized Kadanoff–Baym ansatz (GKBA), and exact diagonalization for small systems using second order many-body perturbation theory. Using a driven two-band Hubbard model, we show that mean-field single particle density matrix trajectories provide a reliable baseline for RTDE across a broad range of interaction strengths and excited-carrier populations. Further, RTDE accurately captures correlation effects in the Green's functions, including long-lived oscillations and revivals that are strongly suppressed by the overdamping inherent to self-consistent KBE schemes. As a consequence, RTDE resolves rich non-equilibrium spectral structure in time-resolved photoemission, such as interaction- and population-dependent quasiparticle splittings and bandgap renormalization, which are largely washed out in self-consistent approaches, yet are present in the exact solutions. Our results demonstrate that RTDE bridges the gap between mean-field propagation and full two-time KBE simulations, retaining favorable linear scaling while capturing essential dynamical correlations relevant for ultrafast spectroscopy. 
\end{abstract}

\maketitle
\def\thefootnote{*}
\footnotetext{These authors contributed equally to this work.}
\def\thefootnote{\arabic{footnote}} 

\section{Introduction}The non-equilibrium Green’s function (NEGF) formalism provides a rigorous and general framework for describing the dynamics of correlation functions of interacting quantum systems driven far from equilibrium. In practice, it captures phenomena such as ultrafast carrier relaxation, exciton formation, and transient charge transfer in solids and nanostructures \cite{Tuovinen_2023,Perfetto_Wu_Stefanucci_2024,Balzer_2018,Bonitz_Balzer_Schlünzen_Rasmussen_Joost_2019,Murakami_2025,Pavlyukh_2022,Pavlyukh_Tuovinen_2025,Perfetto_Wu_Stefanucci_2024}. The NEGF approach directly connects to experimentally accessible information, such as the time-resolved photoemission spectral functions \cite{freericks_theoretical_2009,Blommel_Perfetto_Stefanucci_Vlček_2025,Murakami_2025,Aoki2014,Giannetti_Capone_Fausti_Fabrizio_Parmigiani_Mihailovic_2016,Kemper_2018,GKBA_KBE_bench} and it is thus indispensable in explaining observations but also guiding experiments towards realization of new transient quantum properties.

At the heart of the NEGF formalism lie the Kadanoff–Baym equations (KBEs)\cite{SvL2013}, which yield, in principle, an exact description of correlated non-equilibrium dynamics. In practice, the interactions are captured via systematically improvable self-energy approximations derived via many-body perturbation theory (MBPT)\cite{MartinSchwinger,Keldysh_diagram}. Regardless of the self-energy form, practical non-equilibrium calculations are however hindered by the two-time structure of the KBEs. This is manifested by time-nonlocal memory effects that lead to cubic computational scaling in the number of time steps \cite{nessi,blommel2024thesis}. Reducing this computational overhead has become a major focus of expanding the NEGF applications beyond small models and short times. Practical schemes include adaptive time grids \cite{Blommel2024,meirinhos2022adaptive,Lang_Sachdev_Diehl_2025}, truncations of the memory integrals \cite{Truncation,AcceleratedCollapse}, or hierarchical decomposition techniques that exploit the low-rank nature of the self-energy and Green's functions \cite{KayeComp2021,Blommel2025}. Recent progress has also drawn on ideas from low-rank and tensor-network representations \cite{Murray_Shinaoka_Werner_2024,Środa_Inayoshi_Shinaoka_Werner_2025}. 

A complementary avenue that holds promise for applications in ab-initio simulations is based on reconstruction of the time-nonlocal correlation functions. These approaches lead to an effectively time-linear scaling methods. For instance, the G1-G2 generalized Kadanoff–Baym ansatz (GKBA) \cite{GKBA_original,Hermanns_2012,G1_G2} yields a single-time linearly scaling propagation scheme for the reduced density matrix. The G1-G2 scheme has enabled simulations of long-time carrier and spin dynamics in Hubbard clusters, semiconductor models, and nanoscale devices \cite{Perfetto_Stefanucci_2023,Bonitz_Vorberger_Bethkenhagen_Böhme_Ceperley_Filinov_Gawne_Graziani_Gregori_Hamann_etal_2024,Joost_2025,Lovato_Bonitz_Balzer_Caruso_Joost,Tuovinen_2023,Balzer_2018,Bonitz_Balzer_Schlünzen_Rasmussen_Joost_2019,Perfetto_Wu_Stefanucci_2024,Pavlyukh2024,Kemper_2018}. To obtain time-dependent spectral function properties, one requires not only the density matrix, but the full two-time Green's function. The Real-time Dyson expansion (RTDE) has been introduced to address this \cite{Reeves2024,reeves2025,Blommel_Perfetto_Stefanucci_Vlček_2025}. It reconstructs time-nonlocal information as a Markovian perturbation theory on top of a non-equilibrium mean-field (or time-local) trajectory for the density matrix. Practically, RTDE builds upon the ideas of the G1-G2 scheme by deriving linearly-scaling equations for integrating the Green's function away from the $t=t'$ diagonal. In this framework, dynamical self-energy effects are included during a reconstruction step in the time-off-diagonal direction ($t\neq t'$), but without evaluating the full memory-dependent collision integral. This allows for the capture of correlation-induced features such as transient band renormalization, dissipative dynamics, and spectral broadening\cite{reeves2025,Blommel_Perfetto_Stefanucci_Vlček_2025}. The RTDE thus bridges the gap between mean-field propagation and full KBE simulations, providing a systematically improvable route to non-equilibrium many-body dynamics. 

In this work, we analyze the performance and behavior of such reconstructions and demonstrate that RTDE yields accurate non-equilibrium time-resolved spectral functions for various excitation and coupling regimes. Further, we show that for practical low-order approximation to the self-energy, the Markovian reconstruction remedies some of the shortcomings of the full self-consistent KBE solution, which suffers from significant broadening of the spectral signatures.

\begin{figure}
    \centering
    \includegraphics[width=0.95\linewidth]{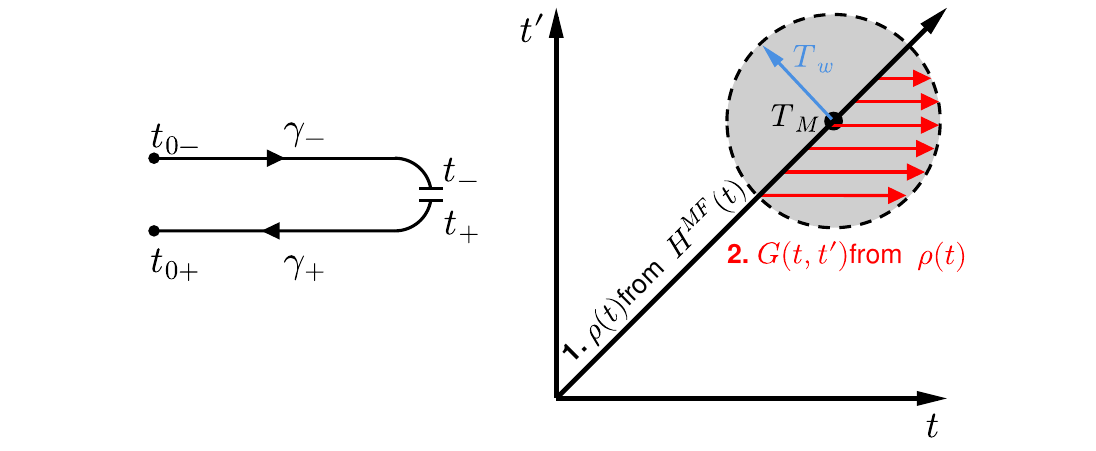}
    \caption{Left: Two-legged Keldysh contour used to calculate non-equilibrium Green's functions. Right: Schematic of RTDE method, which first calculates the mean-field density matrix, which is subsequently used to calculate the off-diagonal Green's function in the region of the photoemission measurement, shown in gray.}
    \label{fig:cartoon}
\end{figure}

\section{Method and Model}
This work compares multiple NEGF methodologies; we will first briefly discuss the formalism and the individual approximations which are a necessary background for the analysis detailed in the next section.
\subsection{NEGF}
The single-particle Green's function captures the evolution of a single quasiparticle probability amplitude for two times $z$ and $z'$ on the Keldysh contour (Fig.~\ref{fig:cartoon}). It is defined as $$G_{ij}(z,z') = \frac{1}{i}\frac{\mathrm{Tr}\left[\mathcal{T}\left\{e^{-i\int_\gamma d\bar{z}\hat{H}(\bar{z})}\hat{d}_i(z)\hat{d}^\dagger_j(z')\right\}\right]}{\mathrm{Tr}\left[\mathcal{T}\left\{e^{-i\int_\gamma d\bar{z}\hat{H}(\bar{z})}\right\}\right]}$$
where $\gamma$ is the two-legged contour, $\mathcal{T}$ is the contour ordering operator, and $\hat{d}^\dagger_j(z')$ creates a particle in orbital $j$ at contour time $z'$.
The lesser and greater Keldysh components are defined by explicitly placing the two contour arguments on either branch, 
\begin{gather}
\begin{aligned}
    G^<(t,t') &= G(t_-,t'_+)\\
    G^>(t,t') &= G(t_+,t'_-)
\end{aligned}
\label{eq:G_keld}
\end{gather}
where $t_\pm$ indicates that the argument lies on the $\gamma_\pm$ branch of the two-legged contour.  It is useful to define the retarded and advanced components as linear combinations of the lesser and greater components
\begin{align}
    G^R(t,t') &= \theta(t-t')[G^>(t,t') - G^<(t,t')]\\
    G^A(t,t') &= \theta(t'-t)[G^<(t,t') - G^>(t,t')]
\end{align}
where $\theta$ is the Heaviside function.
These four Keldysh components obey the following symmetries
\begin{gather}
\begin{aligned}
    G^\gtrless_{ij}(t,t') &= -G^\gtrless_{ji}(t',t)^*\\
    G^R_{ij}(t,t') &= G^A_{ji}(t',t)^*
\end{aligned}
\label{eq:G_symm}
\end{gather}
meaning that the Green's function only needs to be known on the lower half of the $t=t'$ diagonal. 
The single-particle reduced density matrix (henceforth density matrix) is related to the lesser component of the Green's function $\rho(t) = \rho^<(t) = \xi iG^<(t,t)$ where $\xi$ is $+1$ $(-1)$ for bosons (fermions).  In the rest of this manuscript we work with Fermionic systems.  We also will sometimes make use of the greater density matrix $\rho^>(t) = \rho^<(t)-1$.  

\subsection{KBE}
In general, the time-dependent many-body Hamiltonian has the form
\begin{equation}
    \hat{H}(t)=\sum_{i j \sigma} h_{0,i j}(t) \hat{d}_{i \sigma}^{\dagger} \hat{d}_{j \sigma}+\frac{1}{2} \sum_{\substack{i j k l \\ \sigma \sigma^{\prime}}} u_{i j k l}(t) \hat{d}_{i \sigma}^{\dagger} \hat{d}_{j \sigma^{\prime}}^{\dagger} \hat{d}_{k \sigma^{\prime}} \hat{d}_{l \sigma},
\end{equation}
where $h_{0,ij}$ and $u_{ijkl}$ are the one-body and the Coulomb interaction tensor, respectively, both of which can generally be time-dependent. In practical calculations, we employ a model with static two-body term, as detailed in Sec.~\ref{ssec:model}. 

In general, a closed-form solution for the Green's function of such an interacting system does not exist, or it is too computationally expensive to obtain from exact diagonalization, so we must treat the interactions in a perturbative manner.
Information regarding the correlations induced by these electronic interactions is contained within the self-energy $\Sigma(z,z')$, which shares the same Keldysh structure (\ref{eq:G_keld}) and symmetries (\ref{eq:G_symm}) as the Green's function. The self-energy enters into the equations of motion for the Green's function, known as the KBE (written here in matrix form)
\begin{align}
    i\partial_{t}G^R(t,t') &= \epsilon(t)G^R(t,t') \nonumber\\&+ \int_{t'}^{t}d\bar{t} \Sigma^R(t,\bar{t})G^R(\bar{t},t')\label{eq:KBER}\\
    i\partial_t G^<(t,t') &=\epsilon(t)G^<(t,t') + \int_0^td\bar{t}\Sigma^R(t,\bar{t})G^<(\bar{t},t')\nonumber\\&+\int_0^{t'}d\bar{t}\Sigma^<(t,\bar{t})G^A(\bar{t},t')
    \label{eq:KBEL}
\end{align}
where $\epsilon(t) = h_0(t) + \Sigma^{MF}(t)$ is the sum of the one-body Hamiltonian and the mean-field self-energy. The integrals appearing in these equations are generally referred to as collision integrals. The KBE must be solved with the boundary condition for the retarded component on the $t=t'$ diagonal, $G^R_{ij}(t,t) = -i\delta_{ij}$.  

In the perturbative treatment of the interaction term of the many-body Hamiltonian, the self-energy is expressed in a diagrammatic framework.
In this work, we focus on the second-Born approximation given by 
\begin{equation}
    \Sigma_{i j}^\gtrless(t, t^{\prime})=u_{i k l p}(t) u_{q r s j}^{\mathrm{x}}(t')G_{l q}^{\gtrless}(t, t^{\prime}) G_{p r}^{\gtrless}(t, t^{\prime}) G_{s k}^{\lessgtr}(t^{\prime}, t) \label{eq:2B_sigma}
\end{equation}
where we define the ``exchange" interaction tensor $u_{q r s j}^{\mathrm{x}}(t') = -2\xi u_{q r js}(t') -u_{q r s j}(t') $. Note here that we assume spin symmetry, which leads to the factor of two in the first term of the exchange tensor, arising via the sum over spins in the bubble diagram.

\subsection{GKBA}
The direct computation of the Green's function from the KBE is expensive, as it scales cubically with the final integration time. In the evolution of the density matrix, this is avoided via the Generalized Kadanoff-Baym Ansatz (GKBA) \cite{GKBA_original} (here written in matrix form), which reconstructs the time-non-local Green's function using the approximation:
\begin{equation}
    G^{\gtrless,GKBA}(t, t') = -G^{R,MF}(t, t') \rho^{\gtrless}(t')+\rho^{\gtrless}(t) G^{A,MF}(t, t'),
    \label{eq:GKBA}
\end{equation}
where $G^{R,MF}(t,t')$ is the mean-field retarded Green's function whose equation of motion is given by ($\ref{eq:KBER}$) with $\Sigma^R=0$ (similarly for the advanced component).

One can then evaluate the self-energy in Eq.~\ref{eq:KBEL} using the GKBA Green's function instead of the exact, $\Sigma[G]\rightarrow\Sigma[G^{GKBA}]$, which allows for the integro-differential KBE to be rewritten as a pair of coupled ODEs for the density matrix and the two-particle Green's function. 
Their exact form are dependent on the self-energy approximation being used. For our case we will work with the second Born approximation, giving the following EOM
\begin{widetext}
\begin{gather}
\begin{aligned}
    \partial_t \rho(t)&=-i\left[\epsilon(t), \rho(t)]+\xi [I(t)+I^\dagger(t)\right]\\
    I_{i m}(t)&=u_{i k l p}(t) \mathcal{G}_{l p m k}(t)\\
    \partial_t \mathcal{G}_{l p m k}(t)= u_{q r s j}^{\mathrm{x}}(t) & {\left[\rho_{l q}^>(t) \rho_{p r}^>(t) \rho_{s k}^<(t)\rho^<_{j m}(t)-\rho_{l q}^<(t) \rho_{p r}^<(t) \rho_{s k}^>(t)\rho^>_{j m}(t)\right]  } \\
& -i \left[\epsilon_{l x}(t) \mathcal{G}_{x p m k}(t, t^{\prime})+\epsilon_{p x}(t) \mathcal{G}_{l x m k}(t, t^{\prime})-\mathcal{G}_{l p m x}(t, t^{\prime}) \epsilon_{x k}-\mathcal{G}_{l p x k}(t, t^{\prime}) \epsilon_{x m}(t)\right],
\label{eq:G1G2}
\end{aligned}
\end{gather}
\end{widetext}
where $[\cdot,\cdot]$ is the matrix commutator and $I$ denotes the collision integral.

\subsection{RTDE}
The G1-G2 equations are an efficient computational method for obtaining density matrix dynamics which incorporate beyond-mean-field electronic correlations. However, it is clear that we must have access to the Green's function away from $t=t'$ in order to obtain spectral information about the system. In a similar manner as the G1-G2 equations, where we applied the GKBA approximation to the EOM for $\rho(t)$, we can instead apply the GKBA to Eqs.~\ref{eq:KBER} and \ref{eq:KBEL} for the retarded and lesser components directly. Doing so allows us to obtain the RTDE equations for the second Born self-energy which are linear scaling equations for each horizontal line in the $(t,t')$ plane\cite{Reeves2024}, shown as red arrows in Fig.~\ref{fig:cartoon}
\begin{widetext}
\begin{gather}
\begin{aligned}
    \partial_t G^{R/<}(t, t^{\prime})&=-i\left[\epsilon(t) G^{R/<}(t, t^{\prime})+I^{R/<}(t, t^{\prime})\right]\\
    I_{i m}^{R/<}(t, t^{\prime})&=u_{i k l p}(t) \mathcal{F}^{R/<}_{l p m k}(t, t^{\prime})\\
    \partial_t \mathcal{F}^{R/<}_{l p m k}(t, t^{\prime})= u_{q r s j}^{\mathrm{x}}(t) & {\left[\rho_{l q}^>(t) \rho_{p r}^>(t) \rho_{s k}^<(t)-\rho_{l q}^<(t) \rho_{p r}^<(t) \rho_{s k}^>(t)\right] G_{j m}^{R/<}(t, t^{\prime}) } \\
& -i \left[\epsilon_{l x}(t) \mathcal{F}^{R/<}_{x p m k}(t, t^{\prime})+\epsilon_{p x}(t) \mathcal{F}^{R/<}_{l x m k}(t, t^{\prime})-\mathcal{F}^{R/<}_{l p m x}(t, t^{\prime}) \epsilon_{x k}(t)\right].
\label{eq:RTDE}
\end{aligned}
\end{gather}
\end{widetext}
For the retarded component, these equations are to be solved with the boundary conditions on the $t=t'$ diagonal $G_{ij}^R(t',t') = -i\delta_{ij}$ and $\mathcal{F}^R(t',t') = 0$. The boundary condition for the lesser Green's function is $G^<(t',t') = -i\xi \rho^<(t')$, while the initial condition for $\mathcal{F}^<(t',t')$ is ambiguous due to the fact that the collision integral in Eq.~(\ref{eq:KBEL}) is not zero for $t=t'$. For this paper, we choose to make an uncorrelated approximation and set $\mathcal{F}^<(t',t') = 0$, which has been shown to lead to accurate results when compared to exact solutions for small systems\cite{Reeves2024,Blommel_Perfetto_Stefanucci_Vlček_2025}.

To close the RTDE equations of motion, it is necessary to first obtain the density matrix $\rho^<(t)$ that appears in (\ref{eq:RTDE}). We adopt the mean-field trajectory $\rho(t) \approx \rho^{MF}(t)$ and
\begin{equation}
    \partial_t\rho^{MF}(t) = -i[\epsilon(t),\rho^{MF}(t)]_\pm.
    \label{eq:EOM_rho}
\end{equation}
The computational approach thus relies on a two-step procedure outlined in Fig.~\ref{fig:cartoon}, in which we first obtain the time-dependent mean-field density matrix and mean-field Hamiltonian from Eq.~\ref{eq:EOM_rho}, and subsequently compute the time-off-diagonal Green's functions from Eq.~\ref{eq:RTDE} shown with red arrows.  Each of these horizontal integrations is independent, as is easily seen from  Eq.~\ref{eq:RTDE}, allowing for time-parallelized execution of the simulation.

\subsection{Photoemission Spectrum}
In TR-ARPES experiments, a system is first excited via a ``pump" at time $t$ and allowed to evolve for some time before being measured with a probe pulse at $T_M=t+\tau$, where $\tau$ is the delay between pump and probe. This spectroscopy technique returns spectral information about the excited state properties of the states in the system which are occupied. These experiments measure the photoemission spectrum, which can be extracted from the non-equilibrium single-particle Green's function via~\cite{freericks_theoretical_2009} 
\begin{align}
    \mathcal{A}^<&(\omega, T_M)\nonumber\\&=\int dtdt^{\prime} e^{-i \omega(t-t^{\prime})} \mathcal{S}(t-T_M) \mathcal{S}(t^{\prime}-T_M) \operatorname{Tr}\left[G^<(t, t^{\prime})\right].\label{eq:ARPES}
\end{align}
In this expression, the functions $\mathcal{S}(t)$ represent the envelope of the probe pulse used in the experiment. In practice, the duration of these pulses are chosen depending on the trade off between the characteristic energy-resolution required by the measurement and the time-resolution of the dynamics of interest. For this work, we choose this function to be a Gaussian with width $T_w$. Due to the fast decay of these Gaussian envelope functions, we truncate the integration to a finite $t,t'$ range and hence the Green's function is required only in a small circular area around the measurement time $T_M$. This is shown as the shaded region in Fig.~\ref{fig:cartoon}. To obtain the photoemission spectrum from RTDE, we can use the linearly-scaling RTDE equations which allow us to avoid calculating $G(t,t')$ in the unshaded region, as we would have to do if using the full KBE, which reduces the scaling of the method from $T_M^3$ to $T_MT_w^2$.

\begin{figure*}
    \centering
    \includegraphics[width=\linewidth]{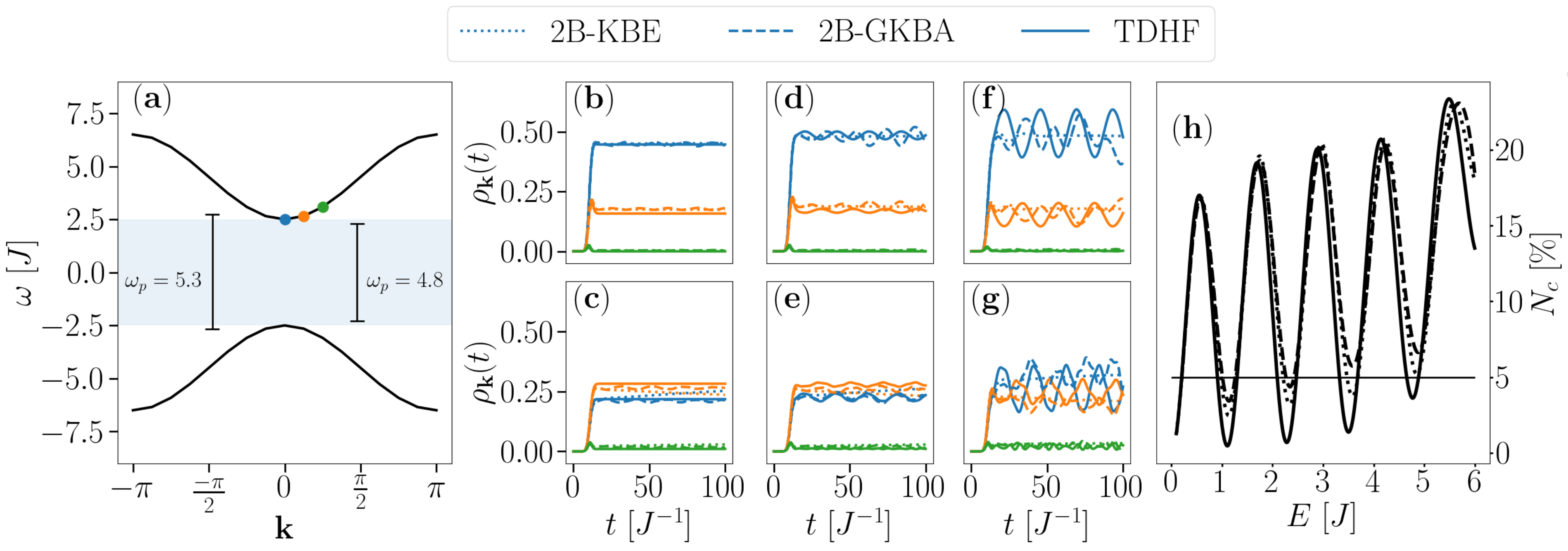}
    \caption{Dynamics of the momentum-resolved photodoped population $\rho_\mathbf{k}(t)$ at different $\mathbf{k}$-points after excitation by a pulse near resonance at $\mathbf{k}=\Gamma$. 
    (a) Equilibrium band structure.
    The shaded area shows the bandgap, and the vertical lines show the above ($\omega_p=5.3$) and below ($\omega_p=4.8$) gap driving.  On the right, the top (bottom) row corresponds to driving the system at $\omega_p=4.8$ $(5.3)$, and the colors correspond to the $\mathbf{k}$-points in (a).
    All results are for intraband interaction $U=1$.
    The first column (b–c) uses interband interaction $V=0$, the second column (d–e) uses $V=0.1$, 
    and the third column (f–g) uses $V=0.5$.  (h) Percentage of electrons excited into the conduction band as the subgap pulse amplitude is varied. The different methods are shown by the full, dashed, and dotted lines (for TDHF, 2B-GKBA, and 2B-KBE).}
    \label{fig:rho_trajectories}
\end{figure*}

\subsection{Model}\label{ssec:model}
In this work, we will use several numerical methods to benchmark the RTDE. We will look at a simple model semiconductor modeled using the two-band Hubbard model with both inter- and intra-band on-site interactions at half-filling
\begin{equation}
    \begin{aligned} \hat{H}=\sum_{\alpha, \beta,i,j,\sigma}h_{i j}^{\alpha \beta}(t) \hat{d}_{i, \sigma}^{\alpha\dagger} \hat{d}_{j, \sigma}^\beta+U\sum_{i, \alpha} \hat{n}_{i \uparrow}^\alpha \hat{n}_{i \downarrow}^\alpha +V \sum_i \hat{n}_{i}^c \hat{n}_{i}^v,
    \label{Eq:Hubb_InterIntra}
\end{aligned}
\end{equation}
where the band indices are represented by Greek characters $\alpha,\beta\in\{c,v\}$, and real-space indices are Latin. The two-body terms are constrained to only contain density-density interactions inside the individual bands and a conduction-valence interband term, with strengths $U$ and $V$. Inspired by the behavior of typical semiconducting materials~\cite{reeves2025}, we consider the behavior for cases when $V<U$, as specified later in the text. The single particle Hamiltonian, $h$, contains intraband nearest-neighbor hopping terms, on-site potentials which set the bandgap, and the interband dipole coupling to an external electric field
\begin{equation}
    h_{i j}^{\alpha \beta}(t) = (-1)^{\delta_{\alpha c}}J\delta_{\langle i,j\rangle}\delta_{\alpha\beta}+\epsilon_\alpha \delta_{ij}\delta_{\alpha\beta}+E(t)\delta_{ij}(1-\delta_{\alpha\beta}).
\end{equation}
In this work, we set the driving field to be sinusoidally oscillating with a Gaussian envelope $E(t)=E\sin[\omega_p(t-t_p)]\exp\left[\frac{-(t-t_p)^2}{2T_p^2}\right]$.  Using this driving field, we can preferentially drive interband transitions at $\mathbf{k}$-points where the band energy difference is resonant with the driving frequency $\omega_p$.
For the rest of this analysis, we set the energy scale relative to the electron hopping amplitude $J=1$, and set the on-site potentials $\epsilon_\alpha$ such that we have the equilibrium bandgap $\Delta=5$ at $\textbf{k}=\Gamma$.  The equilibrium band structure is shown in Fig.~\ref{fig:rho_trajectories}.

\section{Density Matrix Trajectories}
\label{sec:DM}

Before we explore the reconstruction of the time-non-local Green's functions and the spectral functions, we will first investigate the underlying density matrix trajectory. As shown in Fig.~\ref{fig:cartoon}, this is indeed the first step of the RTDE method. Hence, RTDE can be viewed as a ``perturbative'' expansion of non-equilibrium Green's functions around a mean-field (or generally time-local) density matrix trajectory. The accuracy of the density matrix propagation is thus one of the factors determining the RTDE performance. To illustrate this point, we now detail results for a driven system excited with a short-lived electric field pulse at $t_p=10$ with width $T_p=2$. These parameters give us multiple full oscillations of the pulse, allowing us to selectively excite electrons at certain momentum where the two bands are resonant with the pulse. To further expand the regimes investigated, we tune the frequency of the pump field to be slightly off-resonant with the bandgap; namely, we apply a subgap driving frequency of $\omega_p=4.8$ and an above-gap frequency of $\omega_p=5.3$. 

We calculate $\rho(t)$ using three different levels of approximation: Hartree-Fock (HF), 2B using the GKBA approximation (2B-GKBA), and finally the fully self-consistent second Born (2B-KBE). Note that the later approach is the exact time evolution governed by the KBE with the 2B approximation to the self-energy. Results are shown in Fig.~\ref{fig:rho_trajectories} for intraband $U=1$ and several interband interaction parameters between $V=0$ and $V=0.5$. In this example, we tune the amplitude of the pump field $E$ such that we excite approximately $N_c=5\%$ of the electrons into the conduction band, which is a photoexcited density that can be achieved under realistic experimental conditions~\cite{reeves2025}. This tuning process is illustrated in panel (h). We observe that while the driving with larger $E$ amplitudes leads to a higher $N_c$, this trend is overshadowed by large oscillations arising from the strongly nonlinear behavior of the system away from linear response. The use of different approaches leads to slightly different total conduction-band populations, highlighting that while the mean-field and correlated treatments respond similarly to weak excitation, correlations can enhance or suppress absorption in a nontrivial way at stronger driving. 
As the RTDE uses the mean-field trajectory primarily as a reference for constructing the correlated Green’s function, we choose to tune $E$ such that $N_c$ is the same for each approximation.  This provides a consistent baseline for comparing spectral features across approximations.
We note that several different $E$ values can yield the same $N_c$ (as indicated by the intersections of the data with the black line in Fig.~\ref{fig:rho_trajectories}); throughout this work we choose the smallest field amplitude satisfying this condition as larger values of $E$ lead to large intermediate populations which are unrealistic for experimental settings.

We now focus on the comparison of density matrix trajectories computed for various interband couplings. By comparing Fig.~\ref{fig:rho_trajectories} panels b-c (and d-e or f-g) depicting the conduction band density matrix elements, we first observe that for all three methods, the momentum distribution of the photoexcited electrons skews higher in the band for the higher frequency driving, as is expected due to the fact that the higher-frequency driving is resonant at slightly larger momentum. The momentum distribution of the conduction band electrons match well between the three approximations for interband interaction $V=0$.  As we increase $V$ (i.e., from left to right panels b $\to$ d $\to$ f and c $\to$ e $\to$ g), the mean-field and GKBA populations begin to oscillate within the excited region around the $\Gamma$ point. These oscillations are similar in amplitude, with the GKBA data having more structure, stemming from the correlations it contains (in comparison to the HF approximation). In contrast, the oscillations are absent in the self-consistent KBE results obtained with the same self-energy as GKBA. This is consistent with the well-known strong damping of the density matrix response in KBE~\cite{quad_couple1,nessi,GKBA_KBE_bench}. Despite the differences, we note that the TDHF and GKBA results oscillate around the same average value consistent with GKBA.

\begin{figure} 
    \centering
    \includegraphics[width=1\columnwidth]{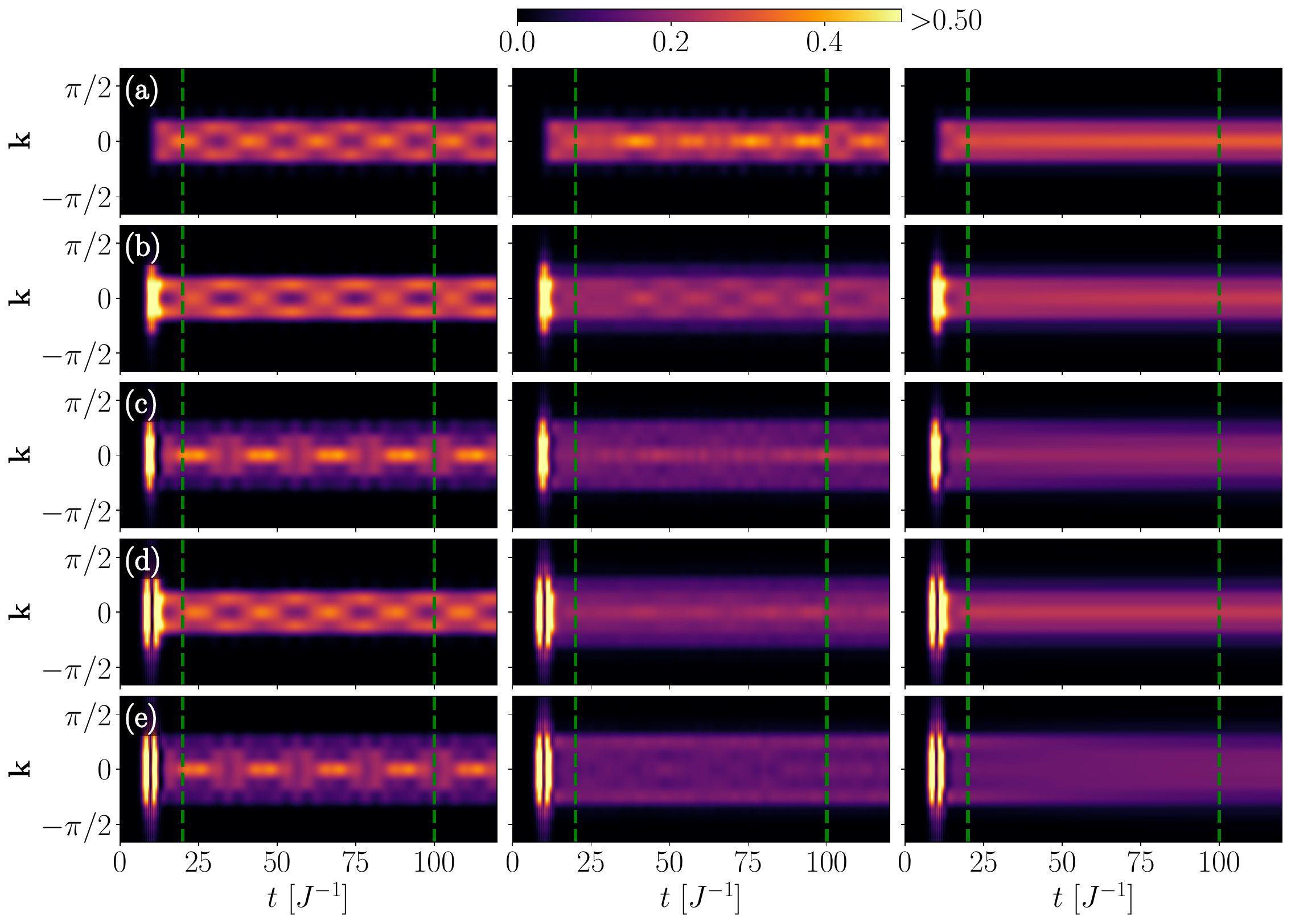}
    \caption{Heatmap plots of $\rho_\mathbf{k}^{HF}(t)$ (left), $\rho_\mathbf{k}^{2B\text{-}GKBA}(t)$ (middle), and $\rho_\mathbf{k}^{2B\text{-}KBE}(t)$ (right) for $V=0.5$, $U=1.0$, and $\omega_p=5.3$.
    The $E$ value of each plot corresponds to the one of the five lowest values that yield $N_c=5\%$ (defined by the intersection of the horizontal line with the curves in Fig.~\ref{fig:rho_trajectories}h).
    Each descending plot corresponds to a higher value of $E$.
    Green dashed lines denote the reconstruction time window used in Sec.~\ref{sec:spect} which has a probe width of 8$\sqrt{2}$}.
    \label{fig:rho_heatmap}
\end{figure}

To further illustrate these density matrix dynamics, we show the evolution of the momentum-resolved populations over time for a representative interaction strength of $V=0.5$, $U=1.0$ in Fig.~\ref{fig:rho_heatmap}. The HF and GKBA calculations display oscillations of carrier populations between the $\Gamma$ point and neighboring $\mathbf{k}$ values following photoexcitation, reflecting coherent intraband motion. The populations in the GKBA and KBE data are typically spread over a wider region of $\mathbf{k}$-space, due to the higher-order correlation effects allowing scattering higher in the band. The oscillation frequencies in GKBA and HF are nearly identical for weak driving, confirming that the mean-field dynamics capture the dominant low-energy transitions and carrier redistribution processes.  Only for strong driving do we see the excitation higher-order modes in the GKBA that are not present in HF, which can be attributed to correlation-induced dephasing. In contrast, the KBE solutions, incorporating all the memory effects, show effectively no coherent intraband population dynamics, which are expected due to the flat nature of the band near the band edge.

In all cases investigated, we saw fast density matrix oscillations compared to the size of the probe window from which the spectral function is computed (explored in the next sections). Due to this, the RTDE calculation are insensitive to the fine details of the density matrix at a particular time  within a single fluctuation. The differences may become imporant for only very short probe windows, which, however, do not yield energetically well-resolved spectra. Further, the close qualitative agreement between TDHF and GKBA supports the use of the mean-field density matrix as a robust input for the RTDE formalism. We remark that the qualitatively same behavior is observed for higher excitations fields $E$, which are however not explored further in this work.

\section{Green's Functions}
\label{sec:GF}
As a next step of the RTDE reconstruction, we focus on the calculations of the time-off-diagonal Green's function in the shaded region shown in Fig.~\ref{fig:cartoon}. Here, we compute horizontal slices of the correlator (red arrows in Fig.~\ref{fig:cartoon}) in the steady-state regime shortly after excitation for a number of different approximation methods: HF, 2B-KBE, 2B-RTDE, and exact diagonalization (ED). 
The comparison allows characterizing the differences among the individual approaches, as well as understanding the physical phenomena captured by each. The results are characteristic and consistent among the various system sizes (including the periodic systems). However, to compare our diagrammatic approaches to the exact solution, we limit the analysis to a two-sites and two-bands model, which can easily be solved exactly via ED.

\begin{figure} 
    \centering
\includegraphics[width=1\columnwidth]{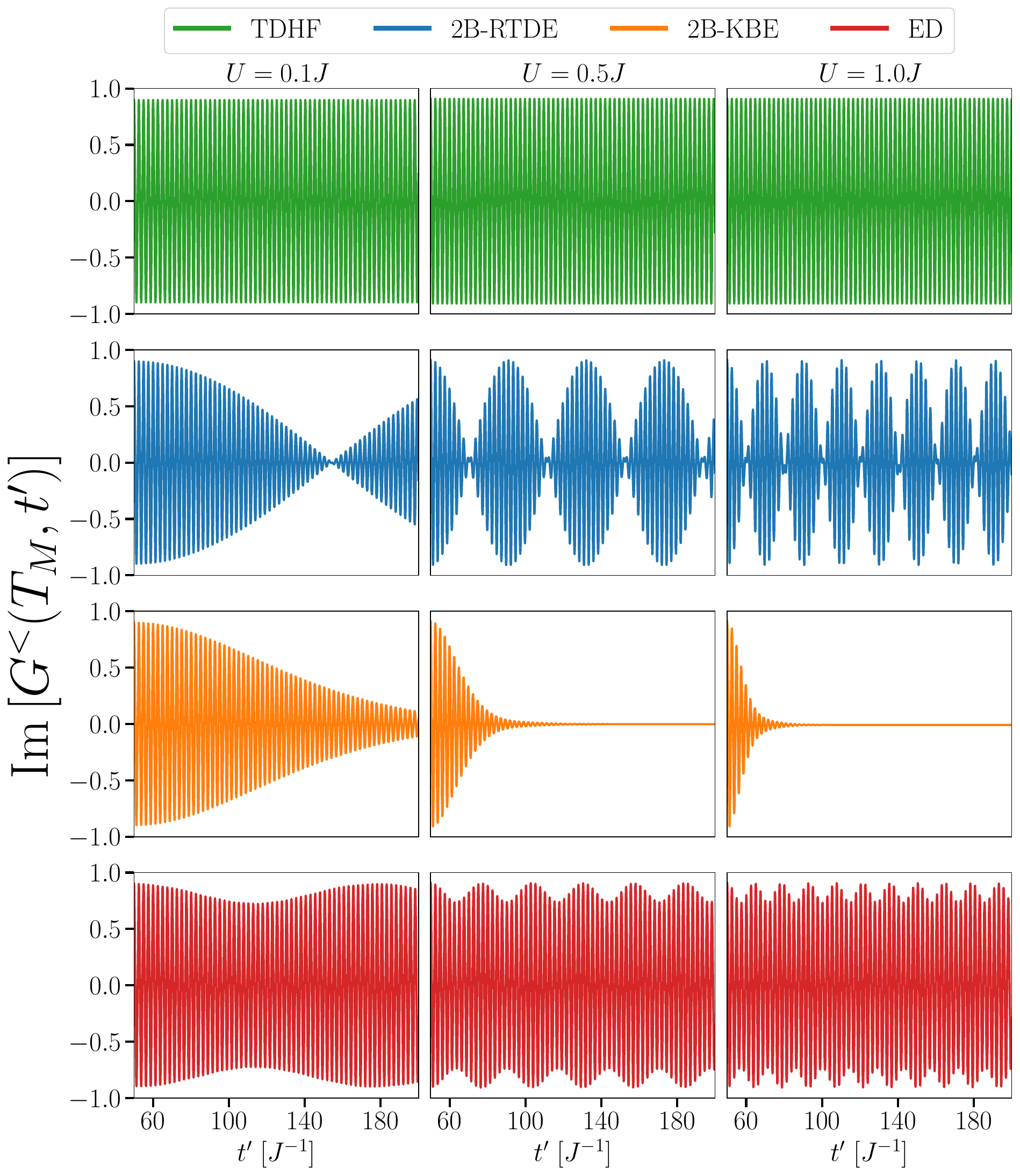}
    \caption{The imaginary component of the lesser Green's function $\mathrm{Im}\left[G^<(T_M,t')\right]$ slice trajectory around the probing time and propagated from $t'=50$ to $t'=200$ using HF, 2B-RTDE, 2B-KBE, and ED with increasing values of $U=\{0.1$, $0.5$, $1.0\}$. Parameters used here are $V=0$, $\omega_p=5.3$, and the driving amplitude $E$ is chosen such that $N_c\approx5\%$.
    }
    \label{fig:GF_trajectories_Uintra}
\end{figure}

We plot the Green's functions for several values of $U$ and $N_c$ in Fig.~\ref{fig:GF_trajectories_Uintra}. Focusing first on the HF results, we see that the Green's functions simply oscillate at a single, fixed frequency without any decay. This is expected, as the HF approximation cannot go beyond the single-particle picture dominated by a transition among a pair of single-particle eigenstates.

In contrast, we see that the ED results (bottom panel in Fig.~\ref{fig:GF_trajectories_Uintra}) exhibit non-equilibrium correlation induced overtones, which are missed by the mean-field results. The exact GF obtained by ED shows that the energy of the low frequency modes (the ``envelope'') is increasing in proportion to the magnitude of the interaction term $U$. Yet, the magnitude of these modes (i.e., the relative scale of the modulation) is constant and independent of $U$.

The solutions obtained with diagrammatic methods (i.e., 2B-KBE and 2B-RTDE) aim to capture higher-order dynamical effects that are missing in the HF results. 
The 2B-RTDE method yields dynamics dominated by a combination of multiple frequencies. While Fig.~\ref{fig:GF_trajectories_Uintra} shows only a single component of the Green's function, this is a representative example (i.e., this is a consistent behavior across all momenta $\mathbf{k}$). Compared to the exact results, the reconstructed GF based via 2B-RTDE differs mainly by exaggerating the strength of the low frequency modulation. Still, the general trend is maintained and the slowly varying envelope qualitatively follows how the ED results change with the increasing interaction strength $U$.

Finally, we note that the 2B-KBE method (representing the formally exact evolution with the second-Born approximation) is plagued by the same problems as those observed for the density matrix propagation. Specifically, the time-off-diagonal Green's function is overdamped. The KBE captures the decay envelope of the ED results only for the first few oscillations, however it is unable to capture the low frequency components associated with the ``revival" and is continually exponentially damped away from the time-diagonal.  As we shall see in the next section, this has a profound impact on the resolution of the spectral functions generated by KBE. We note that the 2B-KBE approximation is based on the fully self-consistent diagrammatic approach, while 2B-RTDE instead evaulates the self-energy as a one-shot correction on top of the mean-field trajectory. Such bare expansions can converge to the correct result in cases where the fully self-consistent approach fails~\cite{Tsuji_Werner_2013}, though the self-consistent expansion has the additional appeal of conserving many physical quantities.

The strong damping in KBE, and the lack thereof in RTDE, illustrates the strength of the latter method over fully self-consistent schemes when used with low-order self-energy approximation, as the lifetimes of spectral features are able to be accurately resolved. Note that the KBE provides the exact solution of the evolution for a given approximate form of the self-energy, i.e., if a an exact self-energy is used, the KBE results shall match the ED evolution. In principle, the exact self-energy would avoid the fast damping observed in Fig.~\ref{fig:GF_trajectories_Uintra}, but we conclude that for low-order self-energy the memory effects lead to damping. In contrast, Markovian RTDE reconstruction does not suffer from these effects. We comment that this failure of the self-consistent methodology (for low order self-energy) can be paralleled with the successes of one-shot $GW$ procedures which are commonly used in materials calculations~\cite{Graml_Zollner_Hernangómez_Pérez_Faria_Junior_Wilhelm_2024,Shao_Lin_Yang_Liu_Da_Jornada_Deslippe_Louie_2016,Shishkin_Kresse_2006,Wen_Abraham_Harsha_Shee_Whaley_Zgid_2024,van2015gw,Körzdörfer_Marom_2012, klein2023one,rasmussen2021towards,schambeck2024solving}. Indeed, this is well documented for a long time\cite{Holm-Barth-scGW,Holm2004-wu} in the context of the $GW$ approximation, in which self-consistent spectral functions  suffer from spurious strong spectral weight transfer between quasiparticle peaks and satellites, and show significant peak broadening. A practical workaround is the combination of the quasiparticle self-consistency (amounting to a Hermitian, but time-local representation)\cite{PhysRevLett.96.226402,PhysRevB.76.165106} with one-shot correction based on the fully dynamical treatment (which is analogous to the application of RTDE).  Hence, it appears that the construction of equilibrium and non-equilibrium spectral functions can be tackled by similar practical remedies.

\begin{figure} 
    \centering
\includegraphics[width=1\columnwidth]{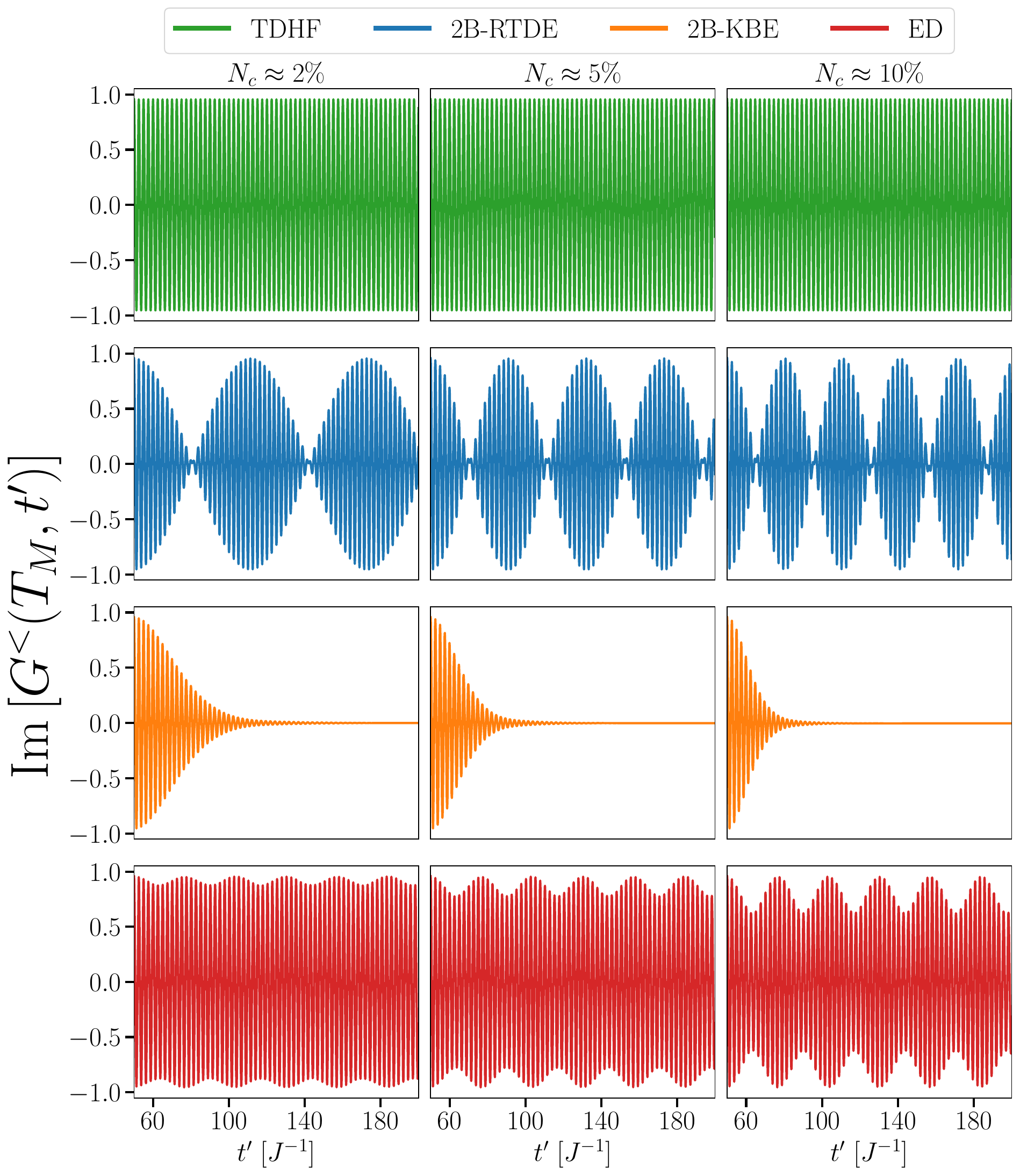}
    \caption{The imaginary component of the lesser Green's function $\mathrm{Im}\left[G^<(T_M,t')\right]$ slice trajectory around the probing time and propagated from $t'=50$ to $t'=200$ using HF, 2B-RTDE, 2B-KBE, and ED with increasing values of $N_c=\{2\%$, $5\%$, $10\%\}$ with fixed $U=0.5$. Parameters used here are $V=0$, $\omega_p=5.3$.
    }
    \label{fig:GF_trajectories_Nc}
\end{figure}

To conclude this section, we now turn to the behavior of the approximations when keeping the interaction strength fixed and varying the percentage of excited charge carriers. These results are shown in Fig.~\ref{fig:GF_trajectories_Nc}, where the excited carrier population changes from $2\%$ to $10\%$. Again, we see that the results for all three approximations match those from Fig.~\ref{fig:GF_trajectories_Uintra} where the intraband interaction $U$ was varied. Inspecting the ED results again reveals a mixture of low and high frequency modes (the latter captured qualitatively by HF). Similar to the dependence on $U$, we see that with increasing $N_c$, the energy of the low frequency modes increases. Interestingly, however, the low frequency envelope becomes more pronounced with increasing $N_c$.

Considering the approximate methods, we see that the 2B-KBE fails to capture revivals, while the 2B-RTDE qualitatively captures the low-frequency envelope oscillations. Yet the performancs of RTDE is not perfect. The low-frequency features become stronger for ED (i.e., the modulation becomes more pronounced), but in 2B-RTDE, these features already dominate the density matrix trajectory. Still, the results show that the 2B-RTDE, in contrast to 2B-KBE, is not agnostic to the excited populations, and can accurately capture strong non-equilibrium correlation effects.

\section{Spectral Functions}
\label{sec:spect}
\begin{figure} 
    \centering
\includegraphics[width=1\columnwidth]{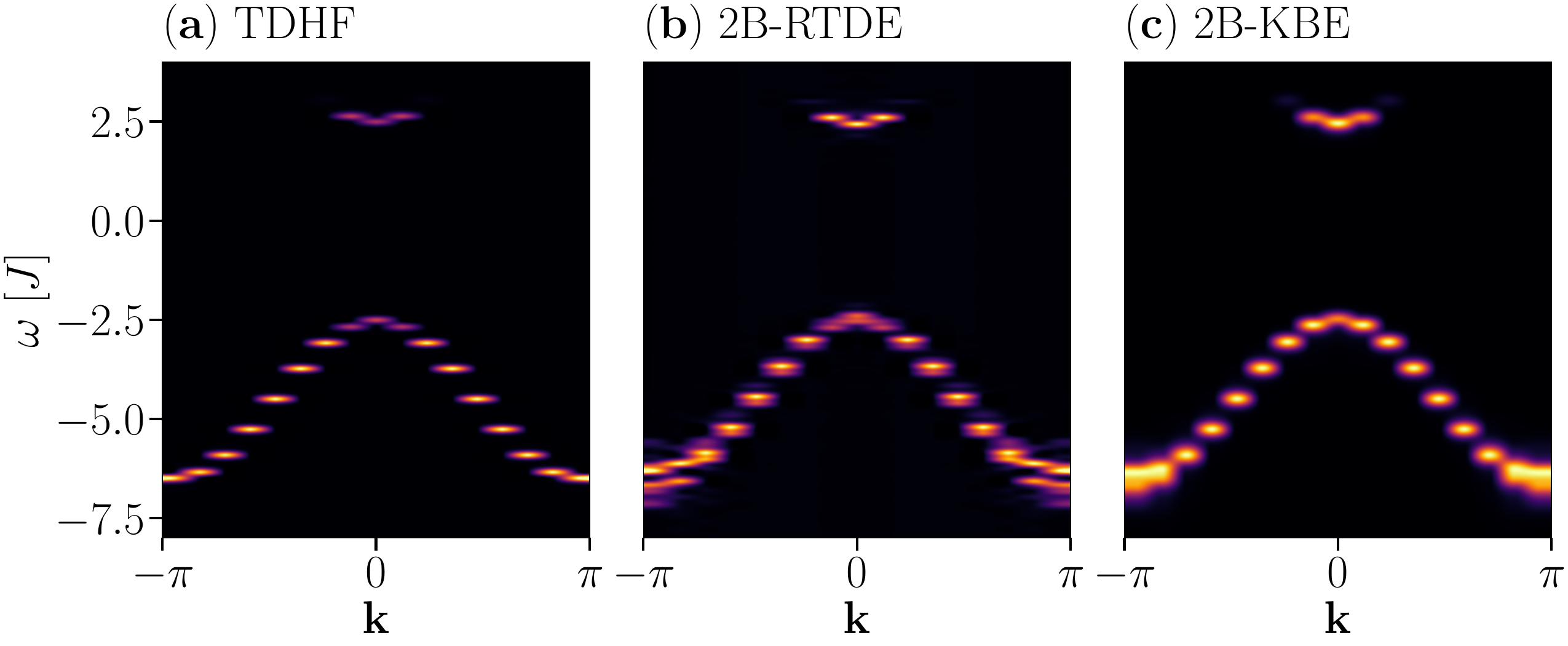}
    \caption{Non-equilibrium time-resolved emission spectrum $\mathcal{A}_{\mathbf{k}}(\omega)$ for the model given in Eq. (\ref{Eq:Hubb_InterIntra}). 
    Panels (a)-(c) shows the spectrum calculated using HF, 2B-RTDE, and 2B-KBE respectively. The model parameters are $U=1.0$, $V=0.5$, $E=0.32$, and $\omega_p=5.3$, with $N_c\approx10\%$.
    The probe width used is $T_w=8\sqrt{2}$, centered at $T_M=60$.
    }
    \label{fig:heatmap_spectrum}
\end{figure}

In the final step, we turn to the performance of various approaches when predicting the spectral functions. This step utilizes the reconstructed Green's functions to compute $\mathcal{A}^< $ from Eq.~\ref{eq:ARPES}. An example photoemission spectrum is shown in Fig.~\ref{fig:heatmap_spectrum}, for a system after the pump pulse with a carrier frequency above the bandgap ($\omega_p=5.3$). The figure compares the results from HF, 2B-RTDE, and 2B-KBE. In contrast to the previous section that used small system to compare to ED, we will focus on spectra obtained from the 16-site model from Sec.~\ref{sec:DM} which is more representative of a realistic system with band dispersion. For all methods, we see that the occupied states appear in spectrum below the Fermi level (set as zero), with amplitude being proportional to the average occupancy. The conduction band is only visible for a small region around $\mathbf{k}=\Gamma$. Results for the below-gap driving are similar, with the resulting spectral weight being transferred lower in the band due to weaker resonance. Comparing the three results we see that the HF spectrum only renormalizes the bands by shifting their energies, otherwise leaving the spectrum unchanged. In contrast, the artificial damping for 2B-KBE discussed in the previous section leads to severe broadening of the spectrum, which is most obvious at the zone edge. This severely limits our ability to study correlation effects and broadening due to the washed-out nature of the results. Continuing the analysis from the previous section, we see that the 2B-RTDE results are more structured than the 2B-KBE results, and we can see how the correlation effects discussed in the previous section lead to a ``splitting" of the bands that is much more pronounced at the zone edge.

\begin{figure*}[t] 
    \centering
\includegraphics[width=0.9\textwidth]{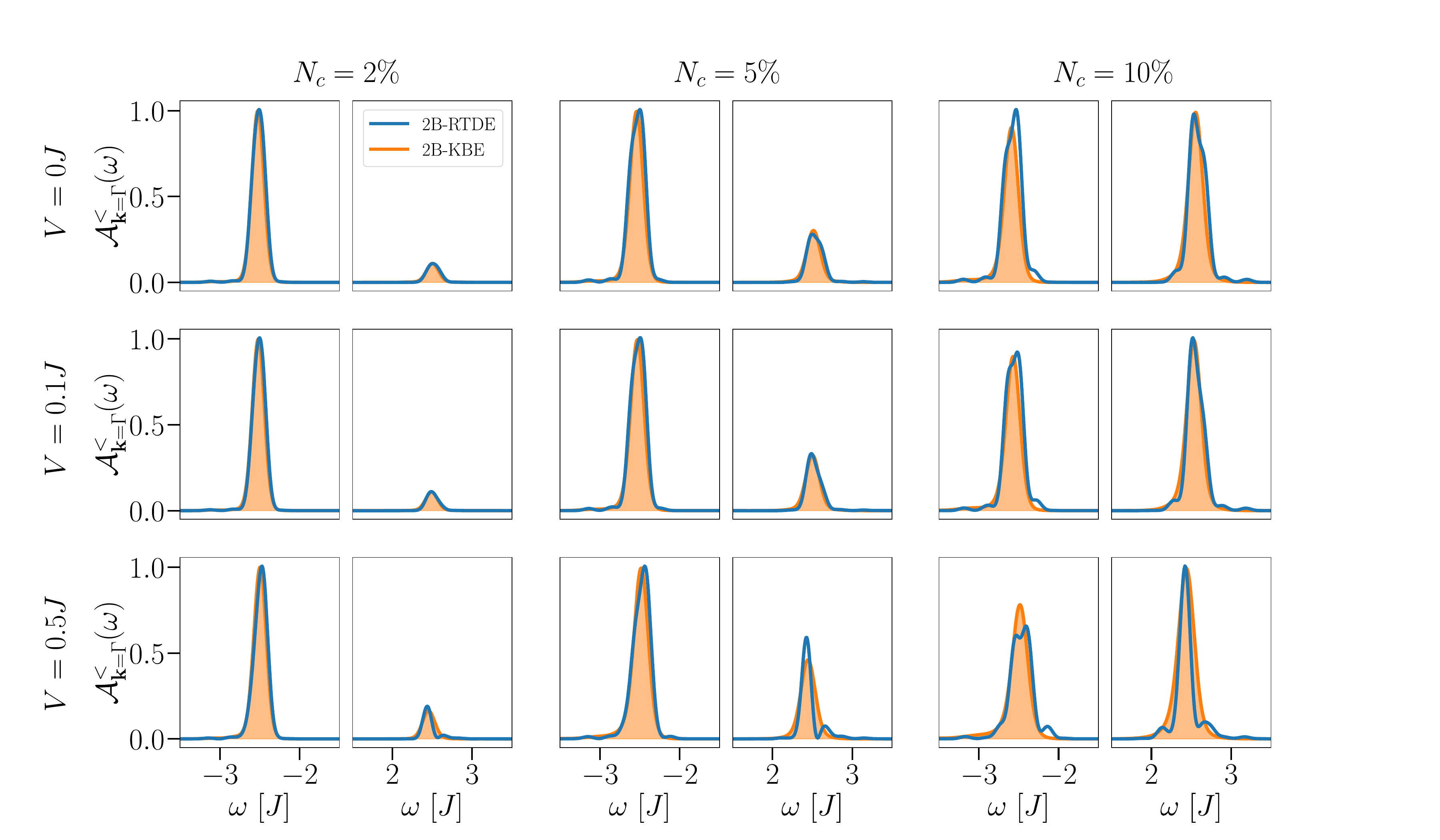}
    \caption{Plots of $\mathcal{A}^{<}_{\mathbf{k}=\Gamma}(\omega)$ 
        for various $N_c$ and $V$ values, comparing 2B-RTDE (blue) and 2B-KBE (orange). The parameters used are $U=1.0$, $\omega_p=5.3$, $V=\{0$, $0.1$, $0.5\}$ with increasing rows, and $N_c=\{2\%$, $5\%$, $10\%\}$ for increasing columns. The odd columns correspond to $\mathcal{A}^{<}_{\mathbf{k}=\Gamma}(\omega)$ below the Fermi level, while the even columns correspond to above the Fermi level.
    }
    \label{fig:spectra_grid}
\end{figure*}

To inspect the results in more detail, we take a slice of this spectrum at the $\Gamma$ point and plot it in Fig.~\ref{fig:spectra_grid}. It is clear that the 2B-KBE results are simply broad Gaussians with no internal structure. The 2B-RTDE results, however, show the appearance of distinct non-equilibrium sub-bands. The number of these visible sub-bands increases from one, at the lowest values of $N_c$ and $V$, to five, for the largest value of $N_c=10\%$, the peak splitting is thus a purely non-equilibrium driven effect. Interestingly, for the largest excited populations, the number of poles does not change as we increase the interband interaction $V$. Furthermore, the distance between these poles increases as the interband interaction is made larger, suggesting an effective repulsion induced by the non-equilibrium excited electron populations. This repulsion increases in strength as the excited populations become larger, indicating stronger correlation effects which go beyond the mean-field. This behavior is somewhat captured by the 2B-KBE results, as the spectrum increases in width as populations increase, however all the information is washed out due to the highly exaggerated damping.

\begin{figure*}[t] 
    \centering
\includegraphics[width=2\columnwidth]{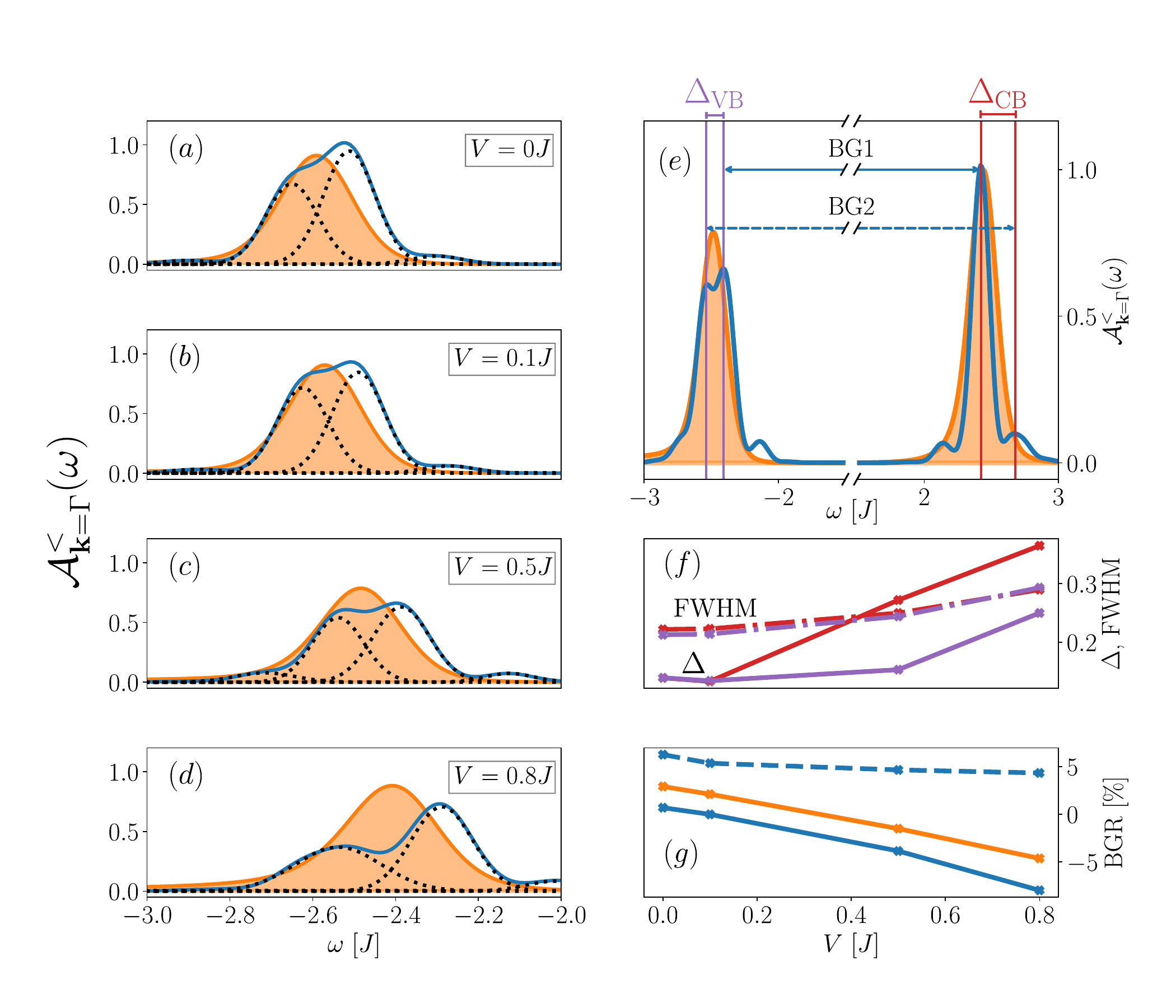}
    \caption{\textit{(a)--(d)} The lesser spectral function $\mathcal{A}_{\mathbf{k}=\Gamma}^{<}(\omega)$ below the Fermi level with increasing $V$ from top to bottom. The black dotted curves show the Gaussian fits used to extract the 2B-RTDE spectral function structure. 
    The orange shaded curve denotes the 2B-KBE spectral function.
    \textit{(e)} Lesser spectral function $\mathcal{A}_{\mathbf{k}=\Gamma}^{<}(\omega)$. We define the peak splitting $\Delta$ as the difference in the fitted peak positions (vertical lines) associated with the two Gaussians composing the main quasiparticle peak. 
    \textit{(f)} Evolution of the 2B-RTDE quasiparticle peak splitting $\Delta$ and the FWHM of 2B-KBE calculations in $\mathcal{A}_{\mathbf{k}}^{<}(\omega)$ as the interband interaction is increased, $V=\{0$, $0.1$, $0.5$, $0.8\}$. 
    The purple and red curves correspond to $\Delta$ and FWHM values extracted below and above the Fermi level, respectively. 
    \textit{(g)} Bandgap renormalization extracted from $\mathcal{A}_{\mathbf{k}}^{<}(\omega)$ of 2B-RTDE and 2B-KBE. For 2B-RTDE, two curves are shown because the main quasiparticle peak is modeled by two Gaussians; the band gap is obtained from the difference between the peak positions above and below the Fermi level. 
    The orange curve shows the corresponding 2B-KBE band gap, extracted from a single-Gaussian fit. In both approaches the band gap decreases with increasing $V$, consistent with larger attraction potential between the conduction electrons and the valence holes. 
    Here we use $U=1$, $\omega_p=5.3$, $T_w=8\sqrt{2}$, $T_M=60$, and $E$ is chosen such that $N_c\approx 10\%$.
    }
    \label{fig:bg_renorm}
\end{figure*}

Figure~\ref{fig:bg_renorm} provides a quantitative analysis of the spectral features that were introduced qualitatively in Figs.~\ref{fig:heatmap_spectrum} and~\ref{fig:spectra_grid} by focusing on the evolution of the dominant quasiparticle peaks of the conduction and valence bands at the $\Gamma$ point as the interband interaction strength $V$ is increased. Panels (a)–(d) show the non-equilibrium 2B-RTDE spectra of the valence band. The valence band splits into five clearly observable peaks, which we fit as the sum of individual Gaussians (each shown as a dotted black line). In contrast, the corresponding 2B-KBE spectra remain characterized by a single, strongly broadened peak. This distinction can be traced directly to the time-domain behavior discussed in Sec.~\ref{sec:GF}: the long-lived oscillations and revivals present in the RTDE Green’s functions give rise to resolvable substructures in the spectrum. However, the excessive damping inherent to the fully self-consistent KBE suppresses these non-Markovian memory effects, effectively merging the same underlying physics into homogeneous lifetime broadening. 

Despite the broadening, the 2B-KBE peaks become more asymmetric with increasing magnitude of $V$, possibly indicating multiple spectral components overshadowed by the broadening effect. Indeed, we expect that as the interaction strength increases, these peaks will both broaden and their separation will increase, reflecting the growing influence of interaction-driven correlations in the photoexcited state. This is confirmed in panel (f), where we plot the separation of the two largest QP peaks in the 2B-RTDE spectrum against the interband interaction strength $V$. We note that the results are extracted from the lesser component of the spectral function, which is not symmetric around the Fermi level ($\omega=0$) and the splitting is distinct for the nominally valence and conduction states as seen in panel (e) This effect is also apparent in the 2B-KBE data, however due to the washed-out nature of the spectrum, we must analyze the full width at the half of the maximum (FWHM) of the single Gaussian state, shown as dashed-dotted lines. We note that the 2B-KBE misses the observed result of 2B-GKBA that the splitting of the conduction band is much larger. Still, for the largest $V$ interaction, we can see that the 2B-KBE spectral function is no longer represented by a regular and symmetric Gaussian (panel d).

The interband interaction drives the splitting of the bands into non-equilibrium QP peaks, but we should also expect that the bandgap decreases due to the presence of stronger interband attraction between the conduction electrons and valence holes~\cite{reeves2025}. This is observed in both our 2B-KBE and 2B-KBE data and to extract the trends, we find the difference between the maximum of the largest underlying QP peaks of the two bands, plotted in (g). Importantly, the RTDE resolves this gap renormalization through well-defined quasiparticle features whose evolution depends on the excited-carrier population, consistent with the population-dependent spectral structures discussed in Fig.~\ref{fig:spectra_grid}. Despite the washed-out nature of the 2B-KBE spectrum, the bandgap renormalization closely tracks that of the 2B-RTDE results. In panel~(g) we also track the distance between the second most prevalent peak in the spectrum. From the latter results, we conclude that the broadening of the KBE is affected by the presence of the secondary peak. Due to the the large broadening of KBE, the secondary peak cannot be well resolved, but the spectral feature is increasingly asymmetric indicating multiple spectral components constituting $\mathcal{A}^<$.

\section{Conclusions}\label{sec:disc_concl}
In this work, we have presented a systematic comparison of the real-time Dyson expansion (RTDE) with established non-equilibrium Green’s function approaches, with the goal of assessing its ability to capture dynamical correlations while retaining favorable computational scaling over the current state of the art methods. 

We showed (Sec.~III) that mean-field density matrix trajectories provide a reliable baseline for RTDE calculations across a range of interaction strengths, driving frequencies, and excited-carrier populations, despite quantitative differences relative to fully self-consistent treatments. Further, we demonstrated (Sec.~IV) that, when propagated on top of these trajectories, the RTDE accurately reproduces correlation-induced features in the time-off-diagonal Green’s functions, including long-lived oscillations and revivals that are absent in self-consistent KBE calculations due to excessive damping. Building on these time-domain results, we established (Sec.~V) that the RTDE yields rich non-equilibrium spectral structure in time-resolved photoemission, resolving interaction- and population-dependent quasiparticle splittings and bandgap renormalization that are washed out in self-consistent approaches. 

Finally, we note the connection between the success of the RTDE approach and methods used for equilibrium spectral functions calculations. RTDE builds perturbatively on top of a mean-field (or generally time-local) reduced density matrix as a one-shot correction. Similarly, first principles approaches that qualitatively capture the spectral features often invoke the scheme in which a one-shot correction is built on otherwise self-consistent but static quasiparticle Hamiltonian. In the context of NEGF simulations, this can be paralleled with adiabatic switching using a time-local approximation followed by the density matrix evolution and RTDE reconstruction.

Taken together, these results show that the RTDE successfully bridges the gap between mean-field propagation and full two-time KBE simulations: it captures essential correlation effects emphasized in the Introduction while avoiding the artificial overdamping and unfavorable scaling that limit traditional methods. This positions the RTDE as a promising and practical framework for interpreting ultrafast spectroscopies and for extending non-equilibrium many-body simulations to larger and more realistic systems.

\section{Acknowledgments}
V.V. acknowledges helpful discussions with Lucia Reining. This material is based upon work supported by the U.S. Department of Energy, Office of Science, Office of Basic Energy Sciences, Computational and Theoretical Chemistry program under Award Number DE-SC0026045. 
This research used resources of the National Energy Research Scientific Computing Center, a DOE Office of Science User Facility supported by the Office of Science of the U.S. Department of Energy under NERSC Award No. BES-ERCAP0035975.

\bibliography{mybib}

\end{document}